\documentclass[aps,preprint,showpacs]{revtex4}
\usepackage{epsfig}
\newcommand{\be}{\begin{equation}}
\newcommand{\ee}{\end{equation}}
\newcommand{\bea}{\begin{eqnarray}}
\newcommand{\eea}{\end{eqnarray}}

\begin{document}
\title{Can re-entrance be observed in force induced transitions?}
\author{Sanjay Kumar}
\affiliation{Department of Physics, Banaras Hindu University,
Varanasi 221 005, India. \\E-mail:yashankit@yahoo.com}

\begin{abstract}
A large conformational change in the reaction co-ordinate and 
the role of the solvent in the formation of base-pairing are combined to 
settle a long standing issue {\it i.e.} prediction of re-entrance in the 
force induced transition of DNA. 
A direct way to observe the re-entrance, {\it i.e} a strand goes to the closed state 
from the open state and again to the open state
with temperature, appears difficult to be achieved in the laboratory. 
An experimental protocol (in direct way) in the constant force ensemble 
is being proposed for the first time that will enable the  observation
of the re-entrance behavior in the force-temperature plane. 
Our exact results for small oligonucleotide that forms a hairpin structure
provide the evidence that re-entrance can be observed.
\end{abstract}

                                                                                
\maketitle

It took almost four decades to realize that unzipping of double
stranded DNA ($dsDNA$) may be achieved with a force applied solely
at one end instead of varying the temperature or pH of the solvent
\cite{bock1, bock2,bhat99,sebas,nelson}. 
This became possible only because of recent advancement in experimental
techniques (optical tweezers, atomic force microscope, magnetic 
tweezers etc.) which provided  important information 
about elastic, structural and functional  properties of  DNA \cite{
busta1, busta2, cluzel, wang, rief1, busta3, busta4}.
Furthermore, these experiments elucidated the mechanism involved in  
crucial processes like transcription and replication \cite{bock1,bock2,
lee,danilow1,danilow2} . 
Consequently by using force as a thermodynamic
variable, many theoretical and numerical studies were done
in establishing a wealth of definite falsifiable predictions
 \cite{bhat99,nelson,trovato,maren2k2,kbs,gk}.
 Among these, a very important role was played by
simple models which are amenable to analytical solution or very accurate
numerical treatments \cite{bhat99,maren2k2,kbs}. 
Such models despite their 
simplicity, have been proved to be quite predictive and provided many 
important information about the cellular processes.
 One of the most 
important predictions is the existence of a re-entrance behavior 
\cite{maren2k2,trovato,kbs} in the force-temperature plane at low
temperature region. Notably now the prediction of re-entrance is not only 
confined to DNA but also for other bio-polymers 
\cite{maren,mks,kg,kijg,naudts,metz}. However, such re-entrance has so far remained
elusive in the experimental studies. 

\begin{figure}[ht]
\begin{center}
\includegraphics[height=2.85in,width=2.85in]{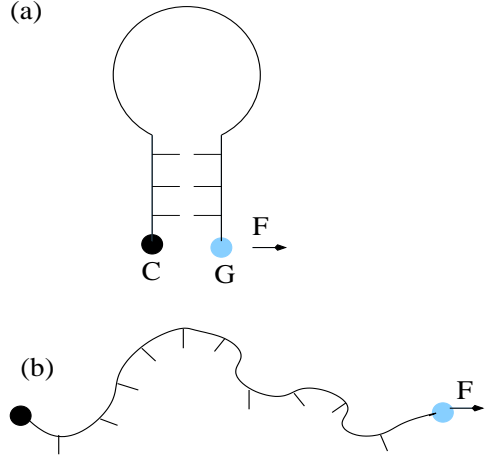}
\caption{(color on line) (a) The schematic representation of a ssDNA 
which forms a hairpin structure. Here bases are on the
links of the strands with short stubs representing
the direction of the hydrogen bonds. 
The small black circle indicates that one end of the 
$ssDNA$ is kept fixed while a force $F$ may be applied at the 
other end shown by the light blue circle. (b) shows the open coil
state induced by a force.}
\label{fig-1}
\end{center}                                                                    
\end{figure}

One of the possible reasons for not observing re-entrance may be  
attributed to the fact that at low temperature
most of the solvents would freeze, precluding  any unzipping experiment.
It is pertinent to mention here that
the temperature ($T^*$) and force ($F^*$) used in these models 
\cite{bhat99,maren2k2} are  not in  real units but in the reduced units
which have been defined as 
\begin{equation}
\frac{1}{T^*} = \frac{\epsilon}{k_BT} 
\end{equation}
\begin{equation}
\frac{F^*x}{T^*} = \frac{Fx}{k_BT}.
\end{equation}

Where $\epsilon$ and $k_B$ are the 
effective base pairing energy and Boltzmann constant respectively. 
Simplyfying  Eqs.(1-2) yield the relations  ( $T = \frac{\epsilon T^* }{k_B}$ 
and $F= \epsilon F^*$) between real units and reduced units.
Moreover, it is also now well
established that the observed re-entrance is due to the ground state entropy. 
Using phenomenological argument \cite{trovato, maren2k2}, it has been shown 
that the critical force for the unzipping is $ F^* = \epsilon + T^* s$ and the 
slope of the phase boundary (zipped-unzipped) at very low temperature 
in the force temperature plane is $\frac {dF}{dT^*} = s$. Here $s$ is 
the entropy 
per base of the dsDNA. 
To observe re-entrance in vitro  $T^*$ as well as  $s$ ( ground state entropy)
should 
be large. However, in case of $dsDNA$ or protein, ground state entropy is not
so large and thus such effect has not been seen.

In a recent controlled experiment, the effect of temperature on single 
stranded DNA ($ssDNA$) has been studied \cite{danilow3}. It appears from 
this experiment 
that at force higher than 12.5 pN, $ssDNA$ attains the stretched state 
and with the rise of temperature, there is an abrupt decrease in extension.
This decrease is not due to formation of hairpins because the solvent
glyoxal (used in the experiment) disrupts the formation of hairpins \cite{
dessinges}. 
Known polymer-based models could not provide the explanation for the 
observed decrease in the extension \cite{danilow3}.  
Kumar and Mishra developed a  lattice model of ssDNA shown in Figure 1 
with suitable  constraints to  show
that this decrease  may be an entropic effect \cite{km}. As a consequence of this, 
the  phase diagram obtained in the force-temperature plane (Figure 2)
for this model also showed a re-entrance at low temperatures.  

These important studies may facilitate to experimental design 
where re-entrance
could be observed \cite{danilow3,dessinges,km}.
The important inferences that can be drawn from the above studies pertain 
to (i) the ideal candidate for 
such experiments should be a single stranded oligonucleotide (a few bases
of both ends are complimentary to each other) which forms a hairpin
structure consisting of a stem ($A-T$ or $C-G$) and 
a loop ($C$ or $G$ for $A-T$ and $A$ or $T$ for $C-G$) \cite{libchaber1, 
libchaber2}.  In this case the reaction co-ordinate (end-to-end 
distance) will exhibit a large conformational change indicating the 
formation/disruption of the hairpin in comparison to $\lambda-$phase DNA. 
Moreover, the ground state entropy of the system will be large
due to the increased contribution from the loop. 
(ii) The reduced temperature of the system may be decreased
or increased by changing the concentration of  glyoxal in the solvent,
keeping the real temperature fixed \cite{danilow3,dessinges}. It is 
pertinent to mention here that 
the solvent glyoxal used in the above experiments reacts with DNA. It
introduces an additional ring to the G-base (to form a tricyclic
compound i.e. glyoxal­dG), thereby sterically preventing
G-C base pair re-annealing \cite{Broude,kasai,Shokri}. Therefore, it may not 
be an ideal 
solvent to use as a denaturating solvent to observe the re-entrance.

\begin{figure}[ht]
\begin{center}
\includegraphics[width=3.7in]{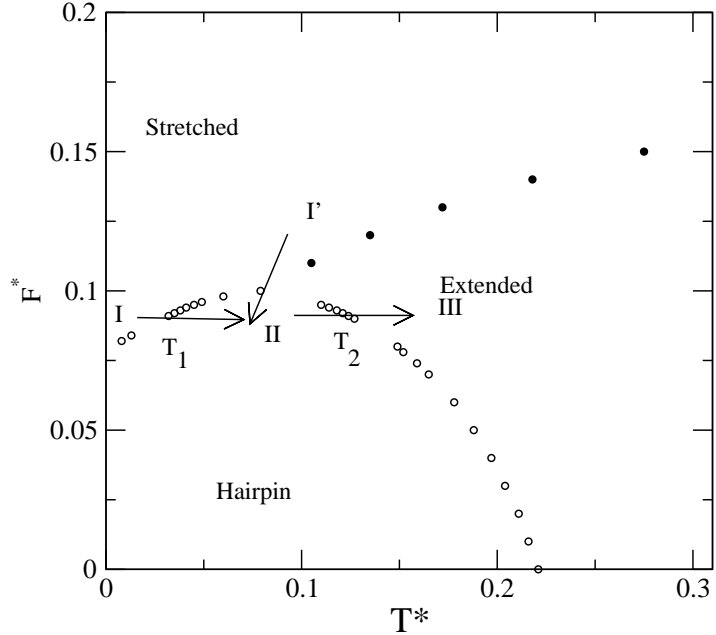}
\caption{The Force-Temperature diagram for DNA hairpin in reduced units.  
A direct way to observe re-entrance is to follow a path $I, II$ and $III$.
Such a path is difficult to achieve in vitro. 
Solid circles
show the cross-over region \cite{km}. By changing the concentration one can
decrease the effective applied force and reduced temperature, keeping the
real temperature and force fixed. Decreses in the reduced temperature
also corresponds to the decrease in the applied force if the system is in
the stretched state (above the upper line  shown by solid circles).
An indirect way to observe re-entrance is 
to follow the path $I', II$ and $III$, in vitro.}

\label{fig-2}
\end{center}
\end{figure}

Thermal melting of DNA occurs at high temperature 
($80-90^o C$). Since
re-entrance has been attributed as a low temperature phenomena,
it is essential that the melting temperature of the system should
be low. There are many solvents which can be used as denaturating 
agents (e.g. 
urea, formamide, formaldehyde, ethidium bromide etc.) to shift the DNA 
melting curve to lower temperatures \cite{McConaughy,Hutton,Bhat,Sadhu}
. In particular, formamide
forms hydrogen bonds with the bases that replace the native ones and 
hence disrupts the base pairing of DNA. It has been seen experimentally
that melting temperature  decreases linearly with the concentration. The 
relation which has has been used quite frequently to determine the melting 
temperature is
\begin{equation}
T_m= 81.5 +16.6 log M +41(xG+xC)-500/L -0.62f
\end{equation}
where $M$ is the molar concentration of monovalent cations, $xG$ and $xC$ are
the mole fractions of $G$ and $C$ in the oligo, $L$ is the shortest strand in 
the stem and $f$ is the molar concentration of formamide \cite{McConaughy, 
Howely}. In fact each $\%$ of change in concentration reduces the
melting temperature by 0.6 degree. The major advantage
of this equation is that it includes the adjustment of salt and formamide.
Moreover, for the Molecular Beacon, the melting temperature is about 
$40^o C$ which can further be lowered by changing the loop length 
\cite{libchaber1,libchaber2}. 
However, for a given loop length,
by adjusting formamide concentration, the melting temperature may be brought
well below the room temperature.

The phase diagram presented in Fig. 2 is in the reduced unit which shows a
peak in force $F^*$ at temperature equal to $T^* \approx 0.09$. To observe 
the re-entrance at a fixed force say$ F^* = 0.09$ , the direct way is to 
follow the path $I, II$ and $III$ by varying the temperature.  
If we take base pairing energy equal to $0.198 10^{-19}$ Joule per base
then melting take place at $313^o K$. Values of $T_1$ 
(open to closed) and $T_2$ (closed to open) are $50^o K$ and $175^o K$
respectively. Such a low temperature and its variation over a large interval 
are difficult to achieve in vitro. In order to observe the re-entrance in vitro, 
we suggest  to
adopt an indirect way ($I'$, $II$ and $III$) which we shall discuss below, 
where by varying the
solvent concentration one can go from the open ($I'$) state
to the closed ($II$) keeping the real temperature fixed.

In the following, 
we propose the experimental protocol, which may be used for the
detection for re-entrance.
The experimental setup used by Danilowicz et al. \cite{danilow3} 
is the starting point but instead of $\lambda-$DNA, we propose to use
Molecular Beacon in presence of formamide that form a hairpin structure 
\cite{libchaber1, libchaber2}. 
From polymer theory, we know that a polymer chain 
will be in either a closed state (hairpin) or 
a swollen state (coil) depending on the temperature \cite{degennes}.  In the 
closed state
(low temperatures) the  average end-to-distance $\langle R \rangle$ will be
nearly equal to zero.  At high temperatures, $\langle R \rangle$ scales 
as $N^\nu$ with $\nu=3/(d+2)$, where $N$ is the length of the chain. 
It is worthwhile to note that in the observed experiment by varying the force,
 polymer chain acquires the conformation of a stretched state.
Hence force induces a new ``stretched state" which is otherwise not accessible
and with the rise of temperature, system attains the extended state.

In presence of formamide, Molecular Beacon will attain the 
stretched state at much lower temperature than $40^o C$ as observed in 
absence of formamide \cite{libchaber1, libchaber2}. Since in this region
(stretched state) stretching force increases with temperatrure (upper line
showen by solid circle in Fig. 2), the required force to keep chain in
stretched state will also be less than $12.5$ pN as seen in the experiment.
If one now reduces  the concentration of the formamide at that 
temperature and force, there will be reannealing of hydrogen bonds 
associated with complementary end bases of Molecular Beacon. 
This corresponds to the increase in the effective base pairing 
energy and thus reduced tempertaure and force of the system will decrease 
in order to keep real temperture and force fixed.
Since the applied force is not sufficient enough to keep Molecular Beacon in 
the stretched state, the reduction in formamide concentration  will drive 
system in to the closed (hairpin) state at the same temperature.
In order to have higher closing rate (probability of forming the
hairpin), we propose to use a stem of $C-G$ and loop made up of $T$. 
Since rate of closing is much higher than the rate of opening 
(probability of opening), this will lead to the formation of base-pairing 
resulting in a hairpin structure \cite{libchaber1,libchaber2}.  If so then there 
will be an abrupt decrease in extension {\it i.e.} reaction co-ordinate 
will approach to zero. Now with the rise of the temperature 
system will 
again attain the open state. It means that in vitro, one would observe 
the system going to the closed state (hairpin) from the open (stretched) 
and again to the open (extended) state i.e $I', II$ and $III$ instead
of $I, II$ and $III$ with rise in the real temperature.

\begin{figure}[ht]
\begin{center}
\vspace{0.3in}
\includegraphics[height=2.4in,width=2.4in]{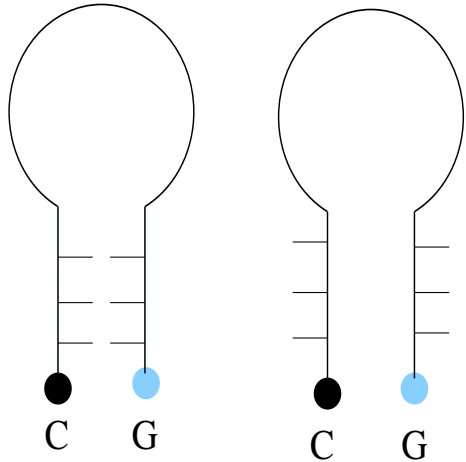}
\caption{Schematic of configurations showing the (a) formation and
(b) non-formation of base pairs.
}
\label{fig-3}
\end{center}
\end{figure}

In order to support the expected outcome of the proposed experiment,
we adopt a more realistic model of DNA \cite{gk,km,kgs}, which may be 
defined in any 
dimension. It was shown that model takes care of important shortcomings
of the known model for ssDNA and incorporates some additional features
such as excluded volume effect, directional nature of hydrogen bond and
the most important configurational entropy which gives rise the 
existence of stretched state. In this model, we consider
a self-avoiding walk of $N$ steps ($N + 1$ vertices) on
a cubic lattice ($3D$). A
step of the walk represents a monomer of the chain. To each monomer a 
base is attached which has a direction associated with it. The first n (=3)
bases of the walk represent nucleotide, $C$ and the last n bases 
the complementary nucleotide  $G$ which forms a stem. The remaining $N - 2n 
= m $ bases constitute a loop of  one type of nucleotides (T) which do not 
participate in pairing. The
repulsion between monomers at short distances (i.e. excluded volume) is 
taken into account by the condition of self-avoidance.  The pairing between 
bases can take place only when the bases at the two ends of the chain
representing the complementary nucleotides approach on the neighboring 
lattice bonds with their directions pointing to each other and perpendicular 
to the lattice bonds they occupy (see Figure 3a). A base can pair at most 
with a complementary base. An energy is gained with formation of each 
base pair. When the two bases representing complementary nucleotides
approach on neighboring lattice bonds but with their directions not 
pointing to each other as shown in Figure 3b, they do not pair and 
therefore there is no gain of energy. 
All possible conformations of ssDNA of N bases mapped by the self-avoiding 
walks  with the constraints specified above and having steps 
N=21 on a cubic lattice (in 3D) have been exactly enumerated. 
The partition function of the system is found from the relation

\begin{figure}[ht]
\begin{center}
\vspace{0.6in}
\includegraphics[height=2.4in,width=2.4in]{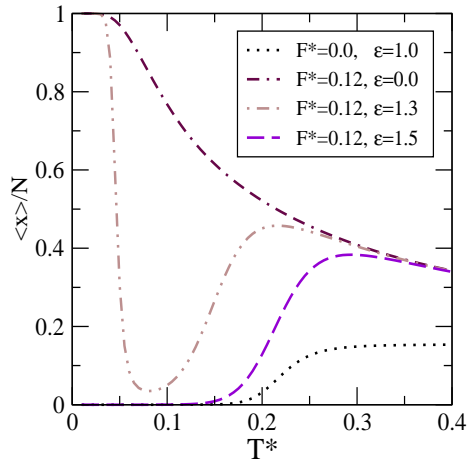}
\caption{(color on line)  Comparison of extension {\it vs} temperature curves
with base pairing ($\epsilon=1.5$) 
and without base-pairing ($\epsilon=0$) 
at constant force 0.12. $\epsilon=1.3$ corresponds to typical concentration 
of formamide, where 
signature of re-entrance can be seen. We also show the melting profile
of hairpin in absence of force ($\epsilon=1.0$ and $F=0$).
}
\label{fig-4}
\end{center}
\end{figure}

\begin{equation}
Z_N = \sum_{p=0}^{n} \sum_{x=0}^{N} C_N(p,x) (e^{-\epsilon/k_B T})^p (e^{-F/k_B T})^x.
\end{equation}

Here, $C_N (p, x)$ is the total number of configurations corresponding to a
walk of $N$ steps with $p$ number of base pairs whose end points are at a distance $x$.
$k_B$, $T$, $\epsilon$ and $F$ are the Boltzmann constant, temperature of the system,
the base pairing energy and the applied force respectively.
Following the method developed in \cite{km,kgs}
we obtained the phase diagram shown in Fig. 2, where we set $\epsilon$ =1 and 
$k_B =1$. 
The stacking energy which arises due to hydrophobicity of bases can be 
incorporated in the description. However, in this work we do not consider 
this interaction explicitly. 

Here we focus ourselves to  study 
the effect of concentration of formamide (formation/disruption of hairpin).
In order to do so we vary the base pairing interaction at constant temperature
and force. $\epsilon =0$ will correspond to oligonucleotide is 
in presence of formamide at high concentration and the the average elongation due to the force
can be found from the relations

\begin{equation}
<x>=\sum x C(m,x) (e^{-\epsilon/k_B T})^p (e^{-F/k_BT})^x.
\end{equation}
The observed decrease in extension (reaction co-ordinate) shown in Fig. 4
with temperature is solely an entropic effect. By setting $k_B = 1$, we get
$F = F^*\epsilon$ and $T=T^*\epsilon$. The increase
in $\epsilon$ means the decrease in formamide concentration and thereby 
decreasing the value of the reduced temperature and the applied force. 
If the applied force is high enough, system will remain in stretched state. 
But if it is in proximity of the upper line, increase in $\epsilon$
(reduction in concentration of formamide) will lead to the base pairing
among end bases in terms of formation of hairpin. This indeed we find in the 
extension-temperature curve shown in Fig. 4, where the 
decrease in extension approaches  zero (showing the formation of hairpin) 
at a particular base-pairing interaction $\epsilon=1.3$. If we increase 
temperature again,
the system attains the open state as predicted by the force-temperature 
diagram  shown in Fig. 2. 

It was shown that the probability distribution curves $P(x)$ with $x$ 
gives important information about the the cellular processes \cite{gk,kg,kgs}
and may
be calculated by using the following relation:  
\begin{equation}
P(x)= \frac{1}{Z_N (m,x)}\sum_{m}
C_N (m,x) (e^{-\epsilon/k_B T})^p (e^{-F/k_BT})^x.
\end{equation}

In Figure  5, we have shown $P(x)$  for different values of temperature
at fixed $F^*=0.12$ and {$\epsilon=1.3$}. The $x$-component of 
the distribution function gives information about the states of oligonucleotide.
At $T^*=0.03$, the peak values is near to $1$, indicating that system is in 
the stretched state. Increase in temperature ($T^*=0.07$) shows the sudden 
shift in peak value near to $0$ which reflects the formation of hairpin. 
Further rise in temperature bring the system from the closed (hairpin) state 
to the open state. This clearly demonstrates that re-entrance can be observed 
at ambient condition.

\begin{figure}[ht]
\begin{center}
\vspace{0.5in}
\includegraphics[height=2.6in,width=2.6in]{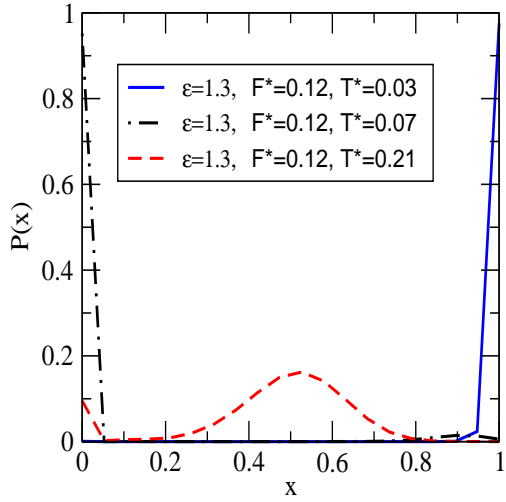}
\caption{(color on line) The probability distribution curve ($P(x)$ 
{\it vs} $x$) for different temperature at fixed force $F$ and base
pairing energy $\epsilon$. These plots clearly show the signature of 
re-entrance. At low temperature ($T^* = 0.03$), oligonucleotide is in the stretched
state. Increase in the temperature ($T^* = 0.07$) brings the system in the closed
(hairpin) state. With further rise in temperature, system attains the open 
state. Here $x$-axis is scaled. 
}
\label{fig-5}
\end{center}
\end{figure}

It may be noted that the present model is a coarse grained,
therefore, quantitative estimate of thermodynamic parameters
are difficult to predict. Moreover, effect of salt concentration, 
pH of the solvent etc are experimental parameters, which have 
been generally ignored in 
the description of coarse grained models. The  model can be
parametrized \cite{eks}, but there is not enough enough data so that  the
link among concentration of formamide, base pairing energy and a 
force can be
established. This requires additional experimental and numerical
works  at this moment to study the effect of force 
in presence of different solvent.

By combining the knowledge accumulated in different context
e.g. kinetic of Molecular Beacon, role of solvent formamide, effect of
temperature in constant force ensemble, formation of hairpin kind of
structure in DNA etc., we proposed  an alternative way that will help
in observing  the re-entrance behavior in bio-polymeric system in 
ambient condition.  The exact solution of the more realistic model of 
ssDNA supported by probability distribution analysis,  renders 
an unequivocal support for the proposed setup.
	
We would like to thank the anonymous referee for bringing out in our notice
about the limited use of glyoxal and for his suggestion to use formamide
as denaturating solvent.
We thank Department of Science and Technology and University Grants
Commission, New Delhi for financial support.
 
 
\end{document}